\documentclass[conference,10pt]{IEEEtran}

\newlength{\figwidth}
\usepackage{xspace}
\usepackage{url}
\usepackage[cmex10]{amsmath}
\usepackage{bbm}
\usepackage{graphicx}
\usepackage{epstopdf}
\usepackage{paralist} 
\usepackage[normalem]{ulem}
\usepackage{fancyref} 
\usepackage{amsmath}
\usepackage{amssymb}
\usepackage[stretch=16,shrink=16,step=4]{microtype}
\usepackage{xcolor}
\definecolor{links}{rgb}{0.3,0,0}   
\definecolor{urls}{rgb}{0,0,0.8}    
\definecolor{cites}{rgb}{0,0,0.6}   
\usepackage[colorlinks,hyperindex,linkcolor=links,citecolor=cites,urlcolor=urls]{hyperref} 
\usepackage{dsfont}
\usepackage{footnote}
\usepackage{balance}
\usepackage{float}
\usepackage{cite}
\usepackage{wasysym}
\usepackage{amssymb}
\let\labelindent\relax
\usepackage[inline]{enumitem}
\usepackage{setspace}
\usepackage{tabularx,booktabs}
\newcolumntype{L}[1]{>{\raggedright\let\newline\\\arraybackslash\hspace{0pt}}m{#1}}
\newcolumntype{C}[1]{>{\centering\let\newline\\\arraybackslash\hspace{0pt}}m{#1}}
\newcolumntype{R}[1]{>{\raggedleft\let\newline\\\arraybackslash\hspace{0pt}}m{#1}}
\usepackage{boldline}
\usepackage{algorithm,algpseudocode}
\usepackage{cuted} 

\algnewcommand{\Initialize}[1]{%
  \State \textbf{Initialize:}
  \Statex \hspace*{\algorithmicindent}\parbox[t]{.8\linewidth}{\raggedright #1}
}

\usepackage{vmr-symbols-vecbold}
\usepackage{standard-macros}
\usepackage{mathbbol}
\usepackage{hyperref}

\DeclareSymbolFontAlphabet{\amsmathbb}{AMSb}%

\newcommand{\lro}[1]{\lefto({#1}\right)}																
\newcommand{\lrho}[1]{\lefto [ {#1} \right ]}																				

\newcommand{\lr}[1]{\left({#1}\right)}																
\newcommand{\lrb}[1]{\left \lbrace {#1} \right \rbrace}															

\safemath{\dopplerspread}{B_D}																								
\safemath{\delayspread}{T_D}																									
\safemath{\nc}{n\sub{c}}																										
\safemath{\nf}{n\sub{f}}																										
\safemath{\efa}{p\sub{sc}}
\safemath{\efb}{p\sub{cs}}
\safemath{\ef}{\epsilon\sub{f}	}
\safemath{\nd}{n\sub{d}}																										
\safemath{\ntx}{n\sub{t}} 																											
\safemath{\nrx}{n\sub{r}}																											
\safemath{\ntxt}{\tilde{n\sub{t}}}																											
\safemath{\cb}{\ensuremath{L}} 																								
\safemath{\cl}{\ensuremath{n}} 																								
\safemath{\txanto}{{\ensuremath{\tilde{m}_t}}} 																		
\safemath{\cs}{M} 																														
\safemath{\idPustm}{\ensuremath{S_{k}}}
\safemath{\error}{\ensuremath{\epsilon}} 																				
\safemath{\eexp}{\ensuremath{\mathcal{E}}} 																			
\safemath{\nsubc}{n\sub{s}}			 																						
\safemath{\nofdm}{n\sub{o}} 																									
\safemath{\bc}{\ensuremath{B_c}} 																							
\safemath{\ts}{\ensuremath{T_s}} 																							
\safemath{\nrb}{\ensuremath{n_{rb}}} 																						
\safemath{\rul}{\ensuremath{\rho\sub{ul}}}
\safemath{\rdl}{\ensuremath{\rho\sub{dl}}}

\safemath{\nres}{\ell}
\safemath{\nr}{n\sub{r}}
\safemath{\maxk}{M^*\lr{\nres, \nsubc, \nofdm, \epsilon, \rho}}
\safemath{\Rmax}{R^*}
\safemath{\Emin}{E\sub{b}^*/N_0}
\safemath{\Eminf}{\frac{E\sub{b}^*}{N_0}}
\safemath{\np}{\ensuremath{n\sub{p}}}
\safemath{\ndf}{\ensuremath{\bar{n}\sub{d}}}
\safemath{\npf}{\ensuremath{\bar{n}\sub{p}}}
\safemath{\code}{\ensuremath{\mathcal{C}}}
\safemath{\err}{\ensuremath{\epsilon}}
\safemath{\rp}{\ensuremath{\rho\sub{p}}}
\safemath{\rd}{\ensuremath{\rho\sub{d}}}
\safemath{\cohtime}{\ensuremath{T\sub{c}}}
\safemath{\cohbw}{\ensuremath{B\sub{c}}}
\safemath{\nmax}{\ensuremath{\ell\sub{m}}}
\safemath{\ntot}{\ensuremath{n\sub{tot}}}
\safemath{\nul}{\ensuremath{n\sub{ul}}}
\safemath{\ndl}{\ensuremath{n\sub{dl}}}

\safemath{\yp}{\ensuremath{\randvecy_{\nu}^{(\text{p})}}}
\safemath{\yd}{\ensuremath{\randvecy_{\nu}^{(\text{d})}}}
\safemath{\ypd}{\ensuremath{\vecy_{\nu}^{(\text{p})}}}
\safemath{\ydd}{\ensuremath{\vecy_{\nu}^{(\text{d})}}}

\safemath{\ypf}{\ensuremath{\bar{\randvecy}_{\nu}^{(\text{p})}}}
\safemath{\ydf}{\ensuremath{\bar{\randvecy}_{\nu}^{(\text{d})}}}
\safemath{\ypdf}{\ensuremath{\bar{\vecy}_{\nu}^{(\text{p})}}}
\safemath{\yddf}{\ensuremath{\bar{\vecy}_{\nu}^{(\text{d})}}}

\safemath{\xp}{\ensuremath{\vecx^{(\text{p})}}}
\safemath{\xd}{\ensuremath{\randvecx_{\nu}^{(\text{d})}}}
\safemath{\xdd}{\ensuremath{\vecx_{\nu}^{(\text{d})}}}

\safemath{\xpf}{\ensuremath{\bar{\vecx}^{(\text{p})}}}
\safemath{\xdf}{\ensuremath{\bar{\randvecx}_{\nu}^{(\text{d})}}}
\safemath{\xddf}{\ensuremath{\bar{\vecx}_{\nu}^{(\text{d})}}}

\safemath{\xdb}{\ensuremath{\overline{\randvecx}^{(\text{d})}}}
\safemath{\Pxd}{\ensuremath{P_{\randvecx^{(\text{d})}}}}

\safemath{\xpbar}{\ensuremath{\overline{\matX}^{(\text{p})}}}
\safemath{\xdbar}{\ensuremath{\overline{\randmatX}^{(\text{d})}}}

\safemath{\xdv}{\ensuremath{\randvecx^{(\text{d})}}}
\safemath{\xdbarv}{\ensuremath{\overline{\randvecx}^{(\text{d})}}}
\safemath{\ydv}{\ensuremath{\randvecy^{(\text{d})}}}

\safemath{\xdr}{\ensuremath{\matX^{(\text{d})}}}

\safemath{\ttx}{\ensuremath{\tau\sub{tx}}}
\safemath{\trx}{\ensuremath{\tau\sub{rx}}}
\safemath{\ack}{\ensuremath{\mathrm{s}}}
\safemath{\nack}{\ensuremath{\mathrm{c}}}

\newcommand{\prob}[1]{\ensuremath{\mathbb{P}\lrho{#1}}}

\safemath{\mI}{\ensuremath{i\lro{\randvecy ; \randvecx}}} 				

\safemath{\randveca}{\bm{A}}
\safemath{\randvecb}{\bm{B}}
\safemath{\randvecc}{\bm{C}}
\safemath{\randvecd}{\bm{D}}
\safemath{\randvece}{\bm{E}}
\safemath{\randvecf}{\bm{F}}
\safemath{\randvecg}{\bm{G}}
\safemath{\randvech}{\bm{H}}
\safemath{\randveci}{\bm{I}}
\safemath{\randvecj}{\bm{J}}
\safemath{\randveck}{\bm{K}}
\safemath{\randvecl}{\bm{L}}
\safemath{\randvecm}{\bm{M}}
\safemath{\randvecn}{\bm{N}}
\safemath{\randveco}{\bm{O}}
\safemath{\randvecp}{\bm{P}}
\safemath{\randvecq}{\bm{Q}}
\safemath{\randvecr}{\bm{R}}
\safemath{\randvecs}{\bm{S}}
\safemath{\randvect}{\bm{T}}
\safemath{\randvecu}{\bm{U}}
\safemath{\randvecv}{\bm{V}}
\safemath{\randvecw}{\bm{W}}
\safemath{\randvecx}{\bm{X}}
\safemath{\randvecy}{\bm{Y}}
\safemath{\randvecz}{\bm{Z}}
\safemath{\randvecphi}{\bm{\Phi}}

\safemath{\randmatA}{\amsmathbb{A}}
\safemath{\randmatB}{\amsmathbb{B}}
\safemath{\randmatC}{\amsmathbb{C}}
\safemath{\randmatD}{\amsmathbb{D}}
\safemath{\randmatE}{\amsmathbb{E}}
\safemath{\randmatF}{\amsmathbb{F}}
\safemath{\randmatG}{\amsmathbb{G}}
\safemath{\randmatH}{\amsmathbb{H}}
\safemath{\randmatI}{\amsmathbb{I}}
\safemath{\randmatJ}{\amsmathbb{J}}
\safemath{\randmatK}{\amsmathbb{K}}
\safemath{\randmatL}{\amsmathbb{L}}
\safemath{\randmatM}{\amsmathbb{M}}
\safemath{\randmatN}{\amsmathbb{N}}
\safemath{\randmatO}{\amsmathbb{O}}
\safemath{\randmatP}{\amsmathbb{P}}
\safemath{\randmatQ}{\amsmathbb{Q}}
\safemath{\randmatR}{\amsmathbb{R}}
\safemath{\randmatS}{\amsmathbb{S}}
\safemath{\randmatT}{\amsmathbb{T}}
\safemath{\randmatU}{\amsmathbb{U}}
\safemath{\randmatV}{\amsmathbb{V}}
\safemath{\randmatW}{\amsmathbb{W}}
\safemath{\randmatX}{\amsmathbb{X}}
\safemath{\randmatY}{\amsmathbb{Y}}
\safemath{\randmatZ}{\amsmathbb{Z}}
\safemath{\randmatSigma}{\mathbb{\Sigma}}
\safemath{\randmatPhi}{\mathbb{\Phi}}
\safemath{\randmatLambda}{\mathbb{\Lambda}}

\safemath{\matSigma}{\bm{\Sigma}}
\safemath{\matPhi}{\bm{\Phi}}
\safemath{\matLambda}{\bm{\Lambda}}

\usepackage{glossaries}
\glsdisablehyper
\newglossaryentry{simo}
{
  name={SIMO},
  description={single-input multiple-output},
  first={\glsentrydesc{simo} (\glsentrytext{simo})}
}

\newglossaryentry{mmse}
{
  name={MMSE},
  description={minimum mean-square error},
  first={\glsentrydesc{mmse} (\glsentrytext{mmse})}
}

\newglossaryentry{na}
{
  name={NA},
  description={normal approximation},
  first={\glsentrydesc{na} (\glsentrytext{na})},
  plural={NAs},
  descriptionplural={normal approximations},
  firstplural={\glsentrydescplural{na} (\glsentryplural{na})}
}

\newglossaryentry{zf}
{
  name={ZF},
  description={zero-forcing},
  first={\glsentrydesc{zf} (\glsentrytext{zf})}
}

\newglossaryentry{rzf}
{
  name={RZF},
  description={regularized zero-forcing},
  first={\glsentrydesc{rzf} (\glsentrytext{rzf})}
}

\newglossaryentry{sir}
{
  name={SIR},
  description={signal-to-interference ratio},
  first={\glsentrydesc{sir} (\glsentrytext{sir})}
}

\newglossaryentry{mse}
{
  name={MSE},
  description={mean-square error},
  first={\glsentrydesc{mse} (\glsentrytext{mse})}
}
\newglossaryentry{mmmse}
{
  name={M-MMSE},
  description={multicell MMSE},
  first={\glsentrydesc{mmmse} (\glsentrytext{mmmse})}
}

\newglossaryentry{ls}
{
  name={LS},
  description={least-squares},
  first={\glsentrydesc{ls} (\glsentrytext{ls})}
}

\newglossaryentry{mr}
{
  name={MR},
  description={maximum ratio},
  first={\glsentrydesc{mr} (\glsentrytext{mr})}
}

\newglossaryentry{ue}
{
  name={UE},
  description={user equipment},
  first={\glsentrydesc{ue} (\glsentrytext{ue})},
  plural={UEs},
  descriptionplural={user equipments},
  firstplural={\glsentrydescplural{ue} (\glsentryplural{ue})}
}

\newglossaryentry{ap}
{
  name={AP},
  description={access point},
  first={\glsentrydesc{ap} (\glsentrytext{ap})},
  plural={APs},
  descriptionplural={access points},
  firstplural={\glsentrydescplural{ap} (\glsentryplural{ap})}
}

\newglossaryentry{cpu}
{
  name={CPU},
  description={central processing unit},
  first={\glsentrydesc{cpu} (\glsentrytext{cpu})},
  plural={CPUs},
  descriptionplural={central processing units},
  firstplural={\glsentrydescplural{cpu} (\glsentryplural{cpu})}
}

\newglossaryentry{bs}
{
  name={BS},
  description={base station},
  first={\glsentrydesc{bs} (\glsentrytext{bs})},
  plural={BS},
  descriptionplural={base stations},
  firstplural={\glsentrydescplural{bs} (\glsentryplural{bs})}
}

\newglossaryentry{ostbc}
{
  name={OSTBC},
  description={orthogonal space-time block code},
  first={\glsentrydesc{ostbc} (\glsentrytext{ostbc})},
  plural={OSTBCs},
  descriptionplural={orthogonal space-time block codes},
  firstplural={\glsentrydescplural{ostbc} (\glsentryplural{ostbc})}
}

\newglossaryentry{biawgn}
{
  name={bi-AWGN},
  description={binary-input additive white Gaussian noise},
  first={\glsentrydesc{biawgn} (\glsentrytext{biawgn})}
}

\newglossaryentry{flnf}
{
  name={FLNF},
  description={fixed-length no-feedback},
  first={\glsentrydesc{flnf} (\glsentrytext{flnf})}
}

\newglossaryentry{crc}
{
  name={CRC},
  description={cyclic redundancy check},
  first={\glsentrydesc{crc} (\glsentrytext{crc})}
}

\newglossaryentry{bsc}
{
  name={BSC},
  description={binary symmetric channel},
  first={\glsentrydesc{bsc} (\glsentrytext{bsc})}
  plural={BSCs},
  descriptionplural={binary symmetric channels},
  firstplural={\glsentrydescplural{bsc} (\glsentryplural{bsc})}
}

\newglossaryentry{bec}
{
  name={BEC},
  description={binary-erasure channel},
  first={\glsentrydesc{bec} (\glsentrytext{bec})}
}
\newglossaryentry{awgn}
{
  name={AWGN},
  description={additive white Gaussian noise},
  first={\glsentrydesc{awgn} (\glsentrytext{awgn})}
}
\newglossaryentry{3gpp}
{
  name={3GPP},
  description={3rd generation partnership project},
  first={\glsentrydesc{3gpp} (\glsentrytext{3gpp})}
}

\newglossaryentry{dl}
{
  name={DL},
  description={downlink},
  first={\glsentrydesc{dl} (\glsentrytext{dl})}
}

\newglossaryentry{LED}
{
  name={LED},
  description={light emitting diode},
  first={\glsentrydesc{LED} (\glsentrytext{LED})},
  plural={LEDs},
  descriptionplural={light emitting diodes},
  firstplural={\glsentrydescplural{LED} (\glsentryplural{LED})}
}

\newglossaryentry{lte}
{
  name={LTE},
  description={long term evolution},
  first={\glsentrydesc{lte} (\glsentrytext{lte})}
}
\newglossaryentry{ofdm}
{
  name={OFDM},
  description={orthogonal frequency-division multiplexing},
  first={\glsentrydesc{ofdm} (\glsentrytext{ofdm})}
}

\newglossaryentry{ul}
{
  name={UL},
  description={uplink},
  first={\glsentrydesc{ul} (\glsentrytext{ul})}
}
\newglossaryentry{rb}
{
  name={RB},
  description={resource block},
  first={\glsentrydesc{rb} (\glsentrytext{rb})},
  plural={RBs},
  descriptionplural={resource blocks},
  firstplural={\glsentrydescplural{rb} (\glsentryplural{rb})}
}
\newglossaryentry{cu}
{
  name={CU},
  description={channel use},
  first={\glsentrydesc{cu} (\glsentrytext{cu})},
  plural={CUs},
  descriptionplural={channel uses},
  firstplural={\glsentrydescplural{cu} (\glsentryplural{cu})}
}
\newglossaryentry{tti}
{
  name={TTI},
  description={transmission time interval},
  first={\glsentrydesc{tti} (\glsentrytext{tti})},
  plural={TTI's},
  descriptionplural={transmission time intervals},
  firstplural={\glsentrydescplural{tti} (\glsentryplural{tti})}
}
\newglossaryentry{iid}
{
  name={i.i.d.},
  description={independent and identically distributed},
  first={\glsentrydesc{iid} (\glsentrytext{iid})}
}
\newglossaryentry{psd}
{
  name={PSD},
  description={power spectral density},
  first={\glsentrydesc{psd} (\glsentrytext{psd})}
}

\newglossaryentry{mimo}
{
  name={MIMO},
  description={multiple-input multiple-output},
  first={\glsentrydesc{mimo} (\glsentrytext{mimo})}
}

\newglossaryentry{bler}
{
  name={BLER},
  description={block error probability},
  first={\glsentrydesc{bler} (\glsentrytext{bler})}
}
\newglossaryentry{mtc}
{
  name={MTC},
  description={machine-type communication},
  first={\glsentrydesc{mtc} (\glsentrytext{mtc})}
}
\newacronym{csi}{CSI}{channel state information}
\newglossaryentry{dvbt}
{
  name={DVB-T},
  description={terrestrial digital video broadcasting},
  first={\glsentrydesc{dvbt} (\glsentrytext{dvbt})}
}
\newglossaryentry{pat}
{
  name={PAT},
  description={pilot-assisted transmission},
  first={\glsentrydesc{pat} (\glsentrytext{pat})}
}
\newglossaryentry{ml}
{
  name={ML},
  description={maximum likelihood},
  first={\glsentrydesc{ml} (\glsentrytext{ml})}
}
\newglossaryentry{dt}
{
  name={DT},
  description={dependence-testing},
  first={\glsentrydesc{dt} (\glsentrytext{dt})}
}
\newglossaryentry{ustm}
{
  name={USTM},
  description={unitary space-time modulation},
  first={\glsentrydesc{ustm} (\glsentrytext{ustm})}
}
\newglossaryentry{siso}
{
  name={SISO},
  description={single-input single-output},
  first={\glsentrydesc{siso} (\glsentrytext{siso})}
}
\newglossaryentry{snn}
{
  name={SNN},
  description={scaled nearest-neighbor},
  first={\glsentrydesc{snn} (\glsentrytext{snn})}
}

\newglossaryentry{snnd}
{
  name={SNND},
  description={scaled nearest-neighbor decoding},
  first={\glsentrydesc{snnd} (\glsentrytext{snnd})}
}
\newglossaryentry{rcus}
{
  name={RCUs},
  description={random-coding union bound with parameter $s$},
  first={\glsentrydesc{rcus} (\glsentrytext{rcus})}
}
\newglossaryentry{osd}
{
  name={OSD},
  description={ordered statistics decoding},
  first={\glsentrydesc{osd} (\glsentrytext{osd})}
}
\newglossaryentry{llr}
{
  name={LLR},
  description={log-likelihood ratio},
  first={\glsentrydesc{llr} (\glsentrytext{llr})}
   plural={LLR's},
  descriptionplural={log-likelihood ratios},
  firstplural={\glsentrydescplural{llr} (\glsentryplural{llr})}
}
\newglossaryentry{qpsk}
{
  name={QPSK},
  description={quaternary phase shift keying},
  first={\glsentrydesc{qpsk} (\glsentrytext{qpsk})}
}
\newglossaryentry{nlos}
{
  name={NLOS},
  description={non-line-of-sight},
  first={\glsentrydesc{nlos} (\glsentrytext{nlos})}
}
\newglossaryentry{los}
{
  name={LOS},
  description={line-of-sight},
  first={\glsentrydesc{los} (\glsentrytext{los})}
}
\newglossaryentry{mgf}
{
  name={MGF},
  description={moment-generating function},
  first={\glsentrydesc{mgf} (\glsentrytext{mgf})}
}
\newglossaryentry{cgf}
{
  name={CGF},
  description={cumulant-generating function},
  first={\glsentrydesc{cgf} (\glsentrytext{cgf})},
  plural={CGFs},
  descriptionplural={cumulant-generating functions},
  firstplural={\glsentrydescplural{cgf} (\glsentryplural{cgf})}
}
\newglossaryentry{pdf}
{
  name={PDF},
  description={probability density function},
  first={\glsentrydesc{pdf} (\glsentrytext{pdf})},
   plural={PDF's},
  descriptionplural={probability density functions},
  firstplural={\glsentrydescplural{pdf} (\glsentryplural{pdf})}
}

\newglossaryentry{ccdf}
{
  name={CCDF},
  description={complementary cumulative distribution function},
  first={\glsentrydesc{ccdf} (\glsentrytext{ccdf})}
}

\newglossaryentry{cdf}
{
  name={CDF},
  description={cumulative distribution function},
  first={\glsentrydesc{cdf} (\glsentrytext{cdf})}
}

\newglossaryentry{gmi}
{
  name={GMI},
  description={generalized mutual information},
  first={\glsentrydesc{gmi} (\glsentrytext{gmi})}
}

\newglossaryentry{arq}
{
  name={ARQ},
  description={automatic repeat request},
  first={\glsentrydesc{arq} (\glsentrytext{arq})}
}

\newglossaryentry{harq}
{
  name={HARQ},
  description={hybrid automatic repeat request},
  first={\glsentrydesc{harq} (\glsentrytext{harq})}
}

\newglossaryentry{fbl}
{
  name={FBL-NF},
  description={fixed blocklength with no feedback},
  first={\glsentrydesc{fbl} (\glsentrytext{fbl})}
}
\newglossaryentry{ssnf}
{
  name={SS-NF},
  description={single-shot no-feedback},
  first={\glsentrydesc{ssnf} (\glsentrytext{ssnf})}
}

\newacronym{urllc}{URLLC}{ultra-reliable low-latency communications}

\newglossaryentry{vlsf}
{
  name={VLSF},
  description={Variable-length stop-feedback},
  first={\glsentrydesc{vlsf} (\glsentrytext{vlsf})}
}

\newglossaryentry{vlnsf}
{
  name={VLNSF},
  description={variable-length noisy stop feedback},
  first={\glsentrydesc{vlnsf} (\glsentrytext{vlnsf})}
}

\newglossaryentry{dmt}
{
  name={DMT},
  description={diversity-multiplexing tradeoff},
  first={\glsentrydesc{dmt} (\glsentrytext{dmt})}
}

\newglossaryentry{dmdt}
{
  name={DMDT},
  description={diversity-multiplexing-delay tradeoff},
  first={\glsentrydesc{dmdt} (\glsentrytext{dmdt})}
}

\newglossaryentry{tddt}
{
  name={TDDT},
  description={throughput-diversity-delay tradeoff},
  first={\glsentrydesc{tddt} (\glsentrytext{tddt})}
}

\newglossaryentry{tdd}
{
  name={TDD},
  description={time-division duplex},
  first={\glsentrydesc{tdd} (\glsentrytext{tdd})}
}

\newglossaryentry{stbc}
{
  name={STBC},
  description={space-time block-codes},
  first={\glsentrydesc{stbc} (\glsentrytext{stbc})},
  plural={STBC's},
  descriptionplural={space-time block-codes},
  firstplural={\glsentrydescplural{stbc} (\glsentryplural{stbc})}
}

\newglossaryentry{harqir}
{
  name={HARQ-IR},
  description={hybrid automatic repeat request with incremental redundancy},
  first={\glsentrydesc{harqir} (\glsentrytext{harqir})}
}

\newglossaryentry{harqcc}
{
  name={HARQ-CC},
  description={hybrid automatic repeat request with chase combining},
  first={\glsentrydesc{harqcc} (\glsentrytext{harqcc})}
}

\newglossaryentry{wp1}
{
  name={w.p.1.},
  description={with probability one},
  first={\glsentrydesc{wp1} (\glsentrytext{wp1})}
}

\newglossaryentry{vlff}
{
  name={VLFF},
  description={variable-length full feedback},
  first={\glsentrydesc{vlff} (\glsentrytext{vlff})}
}

\newglossaryentry{dmc}
{
  name={DMC},
  description={discrete memoryless channel},
  first={\glsentrydesc{dmc} (\glsentrytext{dmc})}
  plural={DMCs},
  descriptionplural={discrete memoryless channels},
  firstplural={\glsentrydescplural{dmc} (\glsentryplural{dmc})}
}

\usepackage{pgfplots}
\usepgfplotslibrary{groupplots}
\usetikzlibrary{pgfplots.groupplots} 

\usepgfplotslibrary{external}
\makeatletter
\def\@IEEEinterspaceratioM{0.265}
\def\@IEEEinterspaceMINratioM{0.1651}
\def\@IEEEinterspaceMAXratioM{0.38}

\def\@IEEEinterspaceratioB{0.31}
\def\@IEEEinterspaceMINratioB{0.19}
\def\@IEEEinterspaceMAXratioB{0.38}
\@IEEEtunefonts
\makeatother
\let\abs\undefined
\newcommand{\abs}[1]{\lvert#1\rvert}		

\let\abs\undefined
\newcommand{\abs}[1]{\lvert#1\rvert}		

 \usepackage{subcaption}
\def\tr{\mathrm{tr}}

\def\diag{\mathrm{diag}}

\newtheorem{theorem}{Theorem}

\def\tr{\mathrm{tr}}

\def\diag{\mathrm{diag}}

\def\tr{\mathrm{tr}}

\def\Htran{\mbox{\tiny $\mathrm{H}$}}
\def\Ttran{\mbox{\tiny $\mathrm{T}$}}
\def\bphiu{\boldsymbol{\phi}} 

 \usepackage[margin=0.9in]{geometry}

\begin{document}
\IEEEoverridecommandlockouts
\IEEEoverridecommandlockouts
\algnewcommand\algorithmicswitch{\textbf{switch}}
\algnewcommand\algorithmicendswitch{\textbf{end switch}}
\algnewcommand\algorithmiccase{\textbf{case}}
\algdef{SE}[SWITCH]{Switch}{EndSwitch}[1]{\algorithmicswitch\ #1\ \algorithmicdo}{\algorithmicend\ \algorithmicswitch }%
\algdef{SE}[CASE]{Case}{EndCase}[1]{\algorithmiccase\ #1}{\algorithmicend\ \algorithmiccase}%
\algtext*{EndSwitch}%
\algtext*{EndCase}%

\title{Cell-free Massive MIMO with Short Packets}
\author{\IEEEauthorblockN{Alejandro Lancho\textsuperscript{*}, Giuseppe Durisi\textsuperscript{*} and Luca Sanguinetti\textsuperscript{\textdagger}}
\IEEEauthorblockA{
  \textsuperscript{*}Chalmers University of Technology, Gothenburg, Sweden\\
  \textsuperscript{\textdagger}University of Pisa, Pisa, Italy\\
  Emails: \{lanchoa,durisi\}@chalmers.se, luca.sanguinetti@unipi.it} \vspace{-0.2cm}
\thanks{This work was partly supported by the Swedish Research Council under grant 2016-03293 and by the Wallenberg AI, autonomous systems, and software program. L. Sanguinetti was in part supported by the Italian Ministry of Education and Research (MIUR) in the framework of the CrossLab project (Departments of Excellence).}
 }
 \maketitle

\begin{abstract}
In this paper, we adapt to cell-free Massive MIMO (multiple-input multiple-output) the finite-blocklength framework introduced by \"Ostman et al. (2020) for the characterization of the packet error probability achievable with Massive MIMO, in the ultra-reliable low-latency communications (URLLC) regime. 
  The framework considered in this paper encompasses a cell-free architecture with imperfect channel-state information, and arbitrary linear signal processing performed at a central-processing unit connected to the access points via fronthaul links. 
  By means of numerical simulations, we show that, to achieve the high reliability requirements in URLLC,  
  MMSE signal processing must be used. 
 Comparisons are also made with both small-cell and Massive MIMO cellular networks. 
  Both require a much larger number of antennas to achieve comparable performance to cell-free Massive MIMO.
\end{abstract}

\smallskip

\section{Introduction}\label{sec:intro}
Massive \gls{mimo} is a key
technology in 5G, owing to its ability to substantially increase the spectral efficiency of cellular networks~\cite{bjornson19-b}. An important challenge in Massive \gls{mimo} is the large pathloss variations and inter-cell interference, in particular for the cell-edge \glspl{ue}~\cite{Sanguinetti20}. An alternative network structure, known as cell-free Massive \gls{mimo}, was recently proposed to overcome this issue~\cite{ngo17-03a,nayebi17-07x}. In this type of network, all the \glspl{ue} in a large coverage area may be  jointly served by multiple distributed \glspl{ap}. The fronthaul connections between the \glspl{ap} and the \gls{cpu}, enable the division of the processing tasks for coherently serving all the active \glspl{ue}.

The aim of this paper is to investigate the design of cell-free Massive \gls{mimo} architectures to support \gls{urllc}---a novel use case in next-generation wireless systems (5G and beyond) aimed at providing connectivity to real-time mission-critical applications, such as remote control of automated factories~\cite{3GPP22.104}.
The low latency required in such applications and the typically small payload contained by the transmitted data packets, make the use of nonasymptotic (i.e., finite-blocklength) information theoretic tools fundamental for the design of such systems~\cite{durisi16-09a}. 

\paragraph*{State of the Art}
In most of the recent literature where the advantages of cell-free Massive \gls{mimo} over traditional architectures are illustrated, ergodic capacity is used as performance metric (see, e.g.,~\cite{bjornson20-1a} and references therein).
Unfortunately, this metric is asymptotic in the blocklength and, hence, not adequate for scenarios in which the packet length is short due to latency constraints.
One recent exception is~\cite{nasir21-04u}, where a conjugate beamforming scheme for cell-free Massive \gls{mimo} architectures is developed on the basis of the so-called \emph{normal approximation}~\cite{polyanskiy10-05a}.
Unfortunately, this approximation, although capturing finite-blocklength effects, tends to loose accuracy at the low error probabilities that are of interest in \gls{urllc}~\cite{ostman20-09b}.
Furthermore, the analysis in~\cite{nasir21-04u} is conducted under the assumption of perfect \gls{csi}.
Hence, it neglects the overhead due to pilot transmission, which is often significant in the short-blocklength regime~\cite{ostman19-02}.
\paragraph*{Contributions}
We illustrate how to use the so-called \gls{rcus} from finite-blocklength length information theory~\cite{martinez11-02a}, to derive both a firm upper bound, and an easy-to compute approximation, based on the saddlepoint method~\cite[Sec. XVI]{feller71-a}, on the \gls{ul} and \gls{dl} error probabilities achievable in a cell-free Massive \gls{mimo} system deployed to support \gls{urllc}.  
To do so, we extend to cell-free Massive-MIMO the finite-blocklength framework derived in~\cite{ostman20-09b} for cellular Massive-MIMO networks. 
  Numerical simulations are used to show that, in a practically relevant automated-factory deployment scenario, cell-free Massive \gls{mimo} with fully centralized processing outperforms cellular Massive \gls{mimo} in terms of the fraction of the coverage area in which \gls{urllc} services can be provided.
\paragraph*{Notation}
Lower-case bold letters are used for vectors and upper-case bold letters are used for matrices.
The circularly-symmetric Gaussian distribution is denoted by $\jpg(0,\sigma^2)$, where $\sigma^2$ denotes the variance.
We use $\Ex{}{\cdot}$ to indicate the expectation operator, and $\prob{\cdot}$ for the probability of a set. The natural logarithm is denoted by $\log(\cdot)$. 
\section{A Finite-Blocklength Upper-Bound on the Error Probability}\label{sec:fbl-intro}
A finite-blocklength information-theoretic upper bound on the error probability achievable in a cell-free Massive \gls{mimo} architecture needs to capture the following elements:
\begin{itemize}
  \item it must allow for linear processing to separate the signals generated by/intended to the different \glspl{ue};
  \item it must allow for pilot-based \gls{csi} acquisition and apply to the scenario where decoding is performed by assuming that the acquired \gls{csi} is exact;
  \item it must apply to a scenario in which the additive noise term includes not only thermal noise, but also channel estimation error and residual multiuser interference after linear processing.
\end{itemize}

A bound satisfying this requirement is the so-called \gls{rcus} given in~\cite{martinez11-02a}.
To introduce this bound, let us consider the following scalar input-output relation:
\begin{equation}\label{eq:simplified_channel}
  v[k] = g q[k] + z[k], \quad k=1,\dots,n.
\end{equation}
Here,  $q[k]$ denotes the $k$th entry of the length-$n$ codeword transmitted by a given user, $v[k]$ is the corresponding received signal after linear processing, $g$ denotes the effective channel after linear processing, which we assume to stay constant over the duration of the packet, and $z[k]$ is the additive noise signal, which also includes the residual multiuser interference after linear processing.

To derive the bound, we shall assume that the receiver does not know $g$ but has access to an estimate $\widehat{g}$ that is treated as perfect.
This estimate may be obtained via pilot transmission, or may simply be based on the knowledge of first-order statistics of $g$.
The first situation is relevant in the \gls{ul} of Massive \gls{mimo}, whereas the second situation typically occurs in the \gls{dl} (see, e.g.,~\cite{bjornson19}).

To determine the transmitted codeword, the decoder performs \gls{snn} decoding, i.e., it seeks the codeword that, after being scaled by the estimated channel gain $\hat{g}$, is closest to the received vector.
Mathematically, the decoder solves the following optimization problem:
\begin{equation}\label{eq:mismatched_snn_decoder}
  \widehat \vecq=\argmin_{\widetilde \vecq\in\mathcal{C}} \vecnorm{\vecv-\widehat{g}\widetilde \vecq}^2.
\end{equation}
Here, $\vecv=[v[1],\dots,v[n]]^{\Ttran}$, the vector $\widehat \vecq$ stands for the codeword chosen by the decoder, and $\mathcal{C}$ denotes the set of length-$n$ codewords.

The \gls{rcus} provides a random coding bound on the error probability $\epsilon=\prob{\widehat \vecq\neq \vecq}$  achieved when the decoder operates according to the rule~\eqref{eq:mismatched_snn_decoder}.
The following theorem provides such a bound for the so-called \emph{Gaussian random ensemble}.

\begin{theorem}[\!\!{\cite[Th.~1]{ostman20-09b}}]\label{thm:rcus}
  Assume that $g\in\mathbb{C}$ and $\widehat{g}\in\mathbb{C}$ in~\eqref{eq:simplified_channel} are random variables drawn according to an arbitrary joint distribution. 
   There exists a coding scheme with $m= 2^b$ codewords of length $n$ operating according to the mismatched \gls{snn} decoding rule~\eqref{eq:mismatched_snn_decoder}, whose error probability $\epsilon$ is upper-bounded by
  \begin{IEEEeqnarray}{lCl} \label{eq:rcus_fading}
    \epsilon &=& \prob{\widehat \vecq\neq \vecq}\nonumber\\
     &\leq& \Ex{g,\widehat{g}}{\prob{\sum_{k=1}^n {\imath_s(q[k],v[k])} \leq \log\frac{m-1}{u} \bigg\given g, \widehat{g}}}
  \end{IEEEeqnarray}
 for all $s>0$. 
 Here, $u$ is a random variable that is uniformly distributed over the interval $[0,1]$ and $\imath_s(q[k],v[k])$ is the so-called \emph{generalized information density}, given by
 \begin{multline}
  \imath_s(q[k],v[k]) = -s \left|{v[k] - \widehat{g} q[k]}\right|^2 \\ + \frac{s\abs{v[k]}^2}{1+s\rho\abs{\widehat{g}}^2} + \log\lro{1+s\rho\abs{\widehat{g}}^2}.
\label{eq:simple_infodens}
\end{multline}
Finally, the average in~\eqref{eq:rcus_fading} is taken over the joint distribution of $g$ and $\widehat{g}$. 
\end{theorem}
\begin{IEEEproof}
See~\cite[App. A]{ostman20-09b}.
\end{IEEEproof}

We refer the interested reader to~\cite{ostman20-09b} for more details about this bound, including its relation to the so-called \emph{generalized mutual information}.
Note that the bound is valid for all values of $s>0$ and can be tightened by performing an optimization over this parameter.

Unfortunately, the bound~\eqref{eq:rcus_fading} is difficult to evaluate numerically.
Indeed, the probability inside the expectation in~\eqref{eq:rcus_fading}  is not known in closed form, and evaluating it accurately for the error-probabilities of interest in \gls{urllc} is time consuming.
One common approach to simplify its evaluation is to invoke the Berry-Esseen central limit theorem~\cite[Ch. XVI.5]{feller71-a} and replace the probability in~\eqref{eq:rcus_fading} with a closed-form approximation that involves the Gaussian $Q(\cdot)$ function and the first two moments of the generalized information density, which are known in closed form.
The resulting approximation, which is usually referred to as the \emph{normal approximation}, has been recently used in~\cite{nasir21-04u} within cell-free Massive \gls{mimo} analyses.
Unfortunately, as shown in~\cite[Fig.~1]{ostman20-09b}, this approximation is accurate only when the rate $R=(\log m)/n$ is close to the expected value of the generalized information density; this is typically not the case for the low error-probabilities of interest in \gls{urllc}.

An alternative approximation, which turns out to be accurate for a much larger range of error-probability values, including the ones of interest in \gls{urllc}, can be obtained using the so-called \emph{saddlepoint method}~\cite[Ch. XVI]{feller71-a}.
The resulting approximation is also in closed form for the setup considered in the present paper.
As a consequence, the saddlepoint approximation has essentially the same computational complexity as
the normal approximation (although its overall expression, given in~\cite[Th.~2]{ostman20-09b}, is arguably more involved).
The saddlepoint approximation depends on the cumulant generating function of the generalized information density and on its first and second derivatives, evaluated at a point that depends on the chosen rate $R$.
In contrast, the normal approximation depends on the mean and the variance of the generalized
information density, i.e., on the value of the first and second-order derivatives of the cumulant
generating function computed at the fixed value $0$.

\section{Cell-free Massive MIMO Network}\label{sec:mimo}

We consider a \emph{fully centralized} network with $L$ \glspl{ap}, each equipped with $M$ antennas, which are geographically distributed over the coverage area. 
The \glspl{ap} serve jointly $K$ single-antenna \glspl{ue}, and are connected via fronthaul links to a \gls{cpu}, which facilitates the \gls{ap} coordination. 
The standard time-division duplexing protocol of cellular Massive \gls{mimo} is used, where the $n$
available channel uses are used for three purposes: 
$\np$ symbols for \gls{ul} pilots; $n_{\rm ul}$ symbols for \gls{ul} data; and $n_{\rm dl}$ symbols for \gls{dl} data. 
The signals received by each \gls{ap} are sent to the \gls{cpu} over the frounthaul links. 
Then, the \gls{cpu} performs both channel estimation and data detection.

The channel between \gls{ap} $l$ and \gls{ue} $i$ is denoted by ${\bf h}_{il} \in \mathbb{C}^{M}$.
We use a correlated Rayleigh fading model where ${\bf h}_{il}\sim \jpg({\bf 0}_M,{\bf R}_{il})$ remains constant for the duration of a codeword transmission. 
The normalized trace $\beta_{il} = \tr({\bf{R}}_{il})/M$ determines the average large-scale fading
between \gls{ap} $l$ and \gls{ue} $i$, while the eigenstructure of ${\bf{R}}_{il}$ describes its spatial channel
correlation~\cite[Sec. 2.2]{bjornson19}.
The collective channel vector ${\bf h}_{i} = [{\bf
h}_{i1}^{\Ttran} \ldots {\bf h}_{iL}^{\Ttran}]^{\Ttran} \in \mathbb{C}^{ML}$ follows a
$\jpg({\bf 0}_{ML}, {\bf{R}}_{i})$ distribution, where ${\bf{R}}_{i} = \diag({\bf{R}}_{i1},\ldots,{\bf{R}}_{iL})$.

\subsection{Pilot Transmission and Channel Estimation}\label{sec:pilots}
The \gls{ul} pilot signature of \gls{ue}~$i$ is denoted by $\bphiu_{i} \in \mathbb{C}^{\np}$ and satisfies $\| \bphiu_{i} \|^2  = \np$.
The elements of $\bphiu_{i}$ are scaled by the square-root of the pilot power $\sqrt{\rho^{\mathrm{ul}}}$ and transmitted over $\np$ channel uses.
This yields the received signal
\begin{IEEEeqnarray}{lCl}
  {\bf Y}_l^{\mathrm{pilot}} = \sqrt{\rho^{\mathrm{ul}}}\sum_{i=1}^K\vech_{il} \bphiu_i^{\Htran} +   {\bf Z}_l^{\mathrm{pilot}}  \label{eq:simo_channel_ul_pilot}
\end{IEEEeqnarray}
where ${\bf Z}_l^{\mathrm{pilot}}\in \mathbb{C}^{M \times \np}$ is noise with i.i.d.\ elements distributed as $\jpg(0,\sigma_{\rm{ul}}^{2})$.  Assuming that the covariance matrices $\{\mathbf{R}_{il}\}$ are known at the \gls{cpu}, the \gls{mmse} estimate of $\vech_{il}$ is \cite[Sec.~3.2]{bjornson19}
 \begin{align} \label{eq:MMSEestimator_h_1}
\widehat \vech_{il}  = \sqrt{\rho^{\mathrm{ul}} \np}{\bf R}_{il} {\bf Q}_{il} ^{-1}  \left({\bf Y}_l^{\mathrm{pilot}}\bphiu_i\right)
 \end{align}
with 
 ${\bf Q}_{il}  = \rho^{\mathrm{ul}} \sum_{i=1}^K{\bf R}_{il} +  \sigma_{\mathrm{ul}}^2  {\bf I}_{M}.$
  The \gls{mmse} estimate $ \widehat \vech_{il} $ and the estimation error $\widetilde \vech_{il}  = \vech_{il}-\widehat \vech_{il}$ are independent random vectors, distributed as $\widehat \vech_{il} \sim \jpg ({\bf 0}, {\bf \Phi}_{il} )$ and $\widetilde{\vech}_{il} \sim \jpg ({\bf 0}, {\bf R}_{il} - {\bf \Phi}_{il} )$, respectively, with ${\bf \Phi}_{il}  = {\rho^{\mathrm{ul}} \np}{\bf R}_{il} {\bf Q}_{il} ^{-1} {\bf R}_{il}$.
To perform coherent processing of the signals at multiple \glspl{ap}, it is necessary to have knowledge of the collective channel ${\bf h}_{i}$, whose estimate is obtained as $\widehat {\bf h}_{i} = [\widehat {\bf h}_{i1}^{\Ttran} \ldots \widehat {\bf h}_{iL}^{\Ttran}]^{\Ttran}$.

\subsection{Uplink Data Transmission}\label{sec:ulDataPhase}
We denote by $x_i^{\rm ul}[k]$ the signal transmitted by \gls{ue} $i$ over channel use $k$.
In the \gls{ul} of a fully centralized operation, each \gls{ap} $l$ acts only as a remote-radio
head, i.e., as a relay that forwards its received baseband signal ${\bf r}_{l}^{\rm{ul}}[k]$ to the \gls{cpu}, which performs detection after linear processing.
Specifically, for $k=1,\ldots,n_{\rm ul}$, the \gls{cpu} computes 
\begin{align}
y_i^{\rm{ul}}[k] = \sum\limits_{l=1}^L{\bf u}_{il}^{\Htran} {\bf r}_{l}^{\rm{ul}}[k] = {\bf u}_{i}^{\Htran} {\bf r}^{\rm{ul}}[k]\label{eq:uplink-CPU-data-estimate}
\end{align}
where ${\bf u}_{i}= [{\bf u}_{i1}^{\Ttran} \, \ldots \, {\bf u}_{iL}^{\Ttran}]^{\Ttran} \in \mathbb{C}^{ML}$ is the centralized linear-combining vector and ${\bf r}^{\rm{ul}}[k]\in \mathbb{C}^{ML}$ is the collective \gls{ul} data signal, given by
\begin{equation} \label{eq:received-data-central2}
{\bf r}^{\rm{ul}}[k] = \begin{bmatrix} {\bf r}_{1}^{\rm{ul}}[k] \\ \vdots \\  {\bf r}_{L}^{\rm{ul}}[k]
\end{bmatrix} = \sum_{i=1}^{K} {\bf h}_{i} x_i^{\rm{ul}}[k] +{\bf z}^{\rm ul}[k]
\end{equation}
with ${\bf z}^{\rm ul}[k]= [{\bf z}_{1}^{{\rm ul}^{\Ttran}}[k] \, \ldots \, {\bf z}_{L}^{{\rm ul}^{\Ttran}}[k]]^{\Ttran} \in \mathbb{C}^{ML}$ being the collective noise vector.

We assume that the \gls{cpu} treats the channel estimate $\widehat{\vech}_i$ as perfect and that the transmitted codeword is drawn from a codebook $\setC^\mathrm{ul}$. The estimated codeword $\widehat{\bf x}_i^{\mathrm{ul}}$ is thus obtained by performing mismatched \gls{snn} decoding with $\widehat{g}= \herm{{\bf u}}_i \widehat{\vech}_i$, i.e.,
\begin{equation}\label{eq:mismatched_snn_decoder-uplink}
  \widehat{\bf x}_i^{\mathrm{ul}}=\argmin_{\widetilde{\bf x}_i^{\mathrm{ul}} \in \setC^\mathrm{ul}} \vecnorm{{\bf y}_i^{\mathrm{ul}}-\widehat{g}\widetilde{\bf x}_i^{\mathrm{ul}}}^2
\end{equation}
with ${\bf y}_i^{\mathrm{ul}} = [y_i^{\mathrm{ul}}[1],\ldots,y_i^{\mathrm{ul}}[\nul]]^{\Ttran}$ and $\widetilde{\bf x}_i^{\mathrm{ul}} = [\widetilde{x}_i^{\mathrm{ul}}[1],\ldots, \widetilde{x}_i^{\mathrm{ul}}[\nul]]^{\Ttran}$.
An upper bound on the error probability then follows by applying~\eqref{eq:rcus_fading}.
This bound is valid for any combining vector. Numerical results will be given for the \gls{mmse} and \gls{mr} combining schemes.

\subsection{Downlink Data Transmission}\label{sec:dlDataPhase}
In the \gls{dl} of a fully centralized network, the \gls{cpu} uses
the \gls{ul} channel estimates to compute the precoding 
vector (by exploiting channel reciprocity) and to transmit the \gls{dl} data signal $x_i^{\rm dl}[k]$ to \gls{ue} $i$ over channel use $k$. Let ${\bf w}_{il} \in \mathbb{C}^M$ denote the precoder that \gls{ap} $l$ assigns to \gls{ue} $i$. In the \gls{dl}, the received signal at \gls{ue} $i$ over channel use $k$, where $k=1,\ldots,n_{\rm dl}$, is
\begin{align}
y_i^{\rm{dl}}[k] &= \sum_{l=1}^{L} {\bf h}_{il}^{\Htran} \sum_{i^\prime=1}^{K}  {\bf w}_{i^\prime l} x_{i^\prime}^{\rm dl}[k] + z_i^{\rm dl}[k] \\&
=  {\bf h}_i^{\Htran}{\bf w}_{i} x_{i}^{\rm dl}[k] + {\bf h}_i^{\Htran}\!\!\!\sum_{i^\prime=1, i^\prime\ne i}^{K}\!\!\!  {\bf w}_{i^\prime} x_{i^\prime}^{\rm dl}[k] + z_i^{\rm dl}[k] \label{eq:Cell-free-DL}
\end{align}
where ${\bf w}_i = [{\bf w}_{i1}^{\Ttran} \, \ldots \, {\bf w}_{iL}^{\Ttran}]^{\Ttran} \in \mathbb{C}^{ML}$ is the collective precoding vector, and $z_i^{\rm dl}[k] \sim \jpg(0,\sigma_{\rm dl}^2)$ is the receiver noise. Without loss of generality, we assume that
\begin{equation} 
{\bf w}_{i} = \sqrt{\rho_i^{\mathrm{dl}}}\bar {\bf w}_{i} 
\end{equation}
where $\vecnorm{ \bar{\bf w}_{i}}^2 = 1$ so that $\rho_i^{\mathrm{dl}}$ can be thought of as the \gls{dl} transmit power.

Since no pilots are transmitted in the \gls{dl}, the \gls{ue} does not know the precoded channel $g = {\bf h}_i^{\Htran}{\bf w}_i$ in \eqref{eq:Cell-free-DL}. 
Instead, we assume that the \gls{ue} has knowledge of its expected value $\Ex{}{{\bf h}_i^{\Htran} {\bf w}_i}$ and uses this quantity to perform mismatched \gls{snn} decoding. 
Specifically, we set $\widehat{g} = \Ex{}{{\bf h}_i^{\Htran} {\bf w}_i}$ and compute the estimated codeword $\widehat{\bf x}_i^{\mathrm{dl}}$ as
\begin{equation}\label{eq:mismatched_snn_decoder-donwlink}
  \widehat{\bf x}_i^{\mathrm{dl}}=\argmin_{\widetilde{\bf x}_i^{\mathrm{dl}} \in \setC^\mathrm{dl}} \vecnorm{{\bf y}_i^{\mathrm{dl}}-\widehat{g}\widetilde{\bf x}_i^{\mathrm{dl}}}^2
\end{equation}
with ${\bf y}_i^{\mathrm{dl}} = [y_i^{\mathrm{dl}}[1],\ldots,y_i^{\mathrm{dl}}[\ndl]]^{\Ttran}$ and $\widetilde{\bf x}_i^{\mathrm{dl}} = [\widetilde{x}_i^{\mathrm{dl}}[1],\ldots, \widetilde{x}_i^{\mathrm{dl}}[\ndl]]^{\Ttran}$. 
Different precoders yield different error probabilities at the \glspl{ue}.
A common heuristic comes from \gls{ul}-\gls{dl} duality~\cite[Sec.~4.3.2]{bjornson19}, which suggests to choose the precoding vectors ${\bf w}_i$ as a function of the combining vectors: ${\bf w}_i= {{\bf u}_i}/{\sqrt{\Ex{}{\vecnorm{{\bf u}_i}^2}}}$.
\section{Numerical Analysis}\label{sec:numericalResult}
We present numerical simulations aimed at investigating the performance of cell-free Massive \gls{mimo} in the \gls{urllc} regime as a function of the number of \glspl{ap}.
Comparisons are made with a cellular Massive \gls{mimo} network to quantify the advantages of the cell-free paradigm. 

\subsection{Network Parameters}
We consider an automated-factory propagation scenario in which the
total coverage area is $150$ m $\times$ $150$ m, and the total number of \glspl{ue} is $K=40$. 
In the cellular setting, we treat the coverage area as one single square cell with a \gls{bs} located in the middle of the cell.
We assume that the \gls{bs} is equipped with a uniform linear array with $LM$ co-located antennas, with half-wavelength spacing.
The \glspl{ue} are independently and uniformly distributed within the cell.  
The cell-free setting is deployed in the same area with the same total number of \glspl{ue} and the same total number of antennas.
Specifically, there are $L$ \glspl{ap} with $M$ antennas each, located at the intersections of a square grid deployed within the coverage area. 
For a fair comparison, we proceed as in \cite{bjornson20-1a} and consider the same propagation model for cell-free Massive \gls{mimo} and cellular Massive \gls{mimo}. 
Hence, we assume that the \glspl{ap} are at $10$~m above the ground. 
As in~\cite{bjornson20-1a}, this vertical distance is only used to impose a minimum distance among \glspl{ap} and \glspl{ue}. 
Specifically, the antennas and the \glspl{ue} are assumed to be located in the same horizontal plane, so that the azimuth angle is sufficient to determine the directivity.  
We use the same \gls{ue} locations and pilot assignments in both cellular and cell-free networks.  

We assume that the scatterers are uniformly distributed in the angular interval $[\varphi_{i} -\Delta, \varphi_{i} + \Delta]$, where  $\varphi_{i}$ is the nominal angle-of-arrival of \gls{ue} $i$, where $i=1,\dots,K$, and $\Delta$ is the angular spread. 
Hence, the $(m_1,m_2)$th element of ${\bf R}_{il}$ is equal to \cite[Sec.~2.6]{bjornson19}
\begin{align}\label{eq:2DChannelModel}
\left[ {\bf R}_{il} \right]_{m_1,m_2} =\frac{\beta_{il}}{2\Delta} \int_{-\Delta}^{\Delta}{ e^{\mathsf{j} \pi(m_1-m_2) \sin(\varphi_{i} + {\bar \varphi}) }}d{\bar \varphi}.
\end{align}
We set $\Delta = 25^\circ$ and let the large-scale fading coefficient, measured in~\dB, be $\beta_{il} \,  = -30.5 - 37.6\log_{10}(d_{il}/1\,\text{m})$,
where $d_{il}$\, is the distance between the \gls{ue} $i$ and the \gls{ap}~$l$. 
The communication takes place over a $20$\,MHz bandwidth with a total receiver noise power of $\sigma_{\mathrm{ul}}^2 = \sigma_{\mathrm{dl}}^2=-96$\,dBm (consisting of thermal noise and a noise figure of $5$\,dB in the receiver hardware) at both the \glspl{ap} and \glspl{ue}. Furthermore, we employ a wrap-around topology as in~\cite[Sec. 4.1.3]{bjornson19}. 
The \gls{ul} and the \gls{dl} transmit powers are equal and given by $\rho^{\mathrm{ul}} = \rho^{\mathrm{dl}} = -10\,\dBm$. 
We assume $n = 300$, $\np=K=40$, $\nul =\ndl=(n-\np)/2$ and $\log_2 m = 160$ information bits, where $m$ is the size of the \gls{ul} and \gls{dl} codebooks $\mathcal{C}^{\mathrm{ul}}$ and $\mathcal{C}^{\mathrm{dl}}$, respectively. 
Note that the assumption $\np=K=40$ implies that an orthogonal pilot sequence is assigned to each \gls{ue} and no pilot contamination occurs.
As pointed out in~\cite{ostman20-09b}, this is crucial to achieve the reliability levels required in \gls{urllc}.

\subsection{Performance Analysis}

\begin{figure}[t!]
    \centering
    \includegraphics[width=\columnwidth]{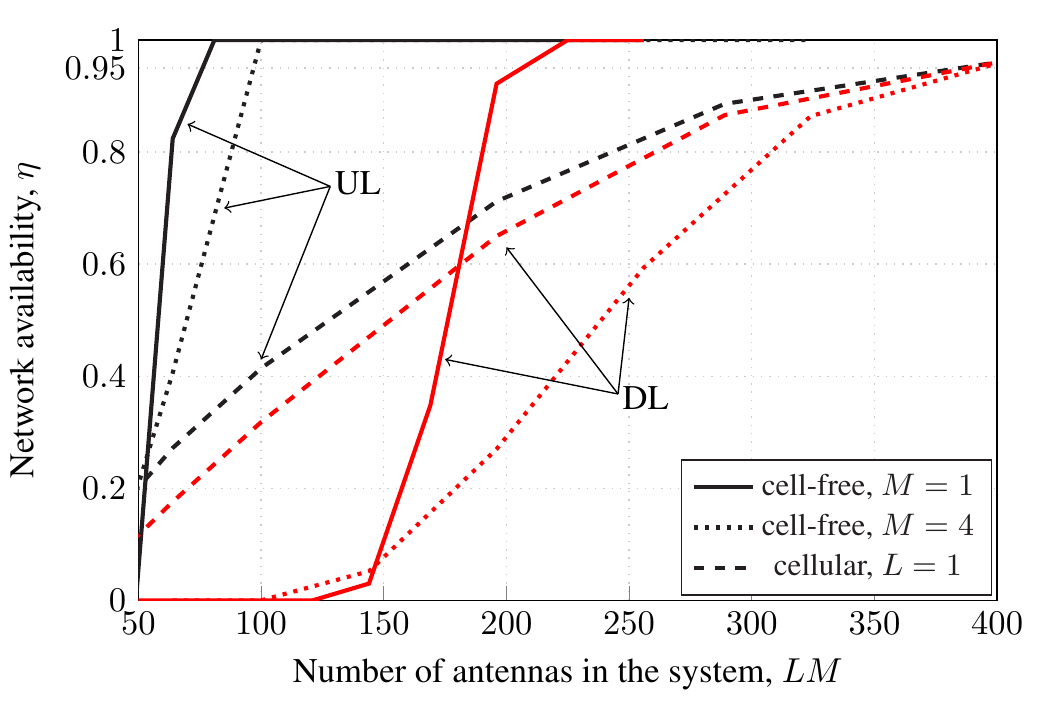}
      \caption{Network availability for $\epsilon\sub{target} = 10^{-5}$ with \gls{mmse} combining/precoding as a function of the total number of antennas in the system $LM$.  
  Here, $K=40$, $\np=40$, $\Delta=25^\circ$, the scenario size is $150\times 150$ m, $\rho^{\mathrm{ul}} = \rho^{\mathrm{dl}}=-10 \dBm$, $\log_2 m = 160$, and $n=300$.}
      \label{fig:avail_vs_antennas_MMSE}
\end{figure}

\begin{figure}[t!]
  \centering
  \includegraphics[width=\columnwidth]{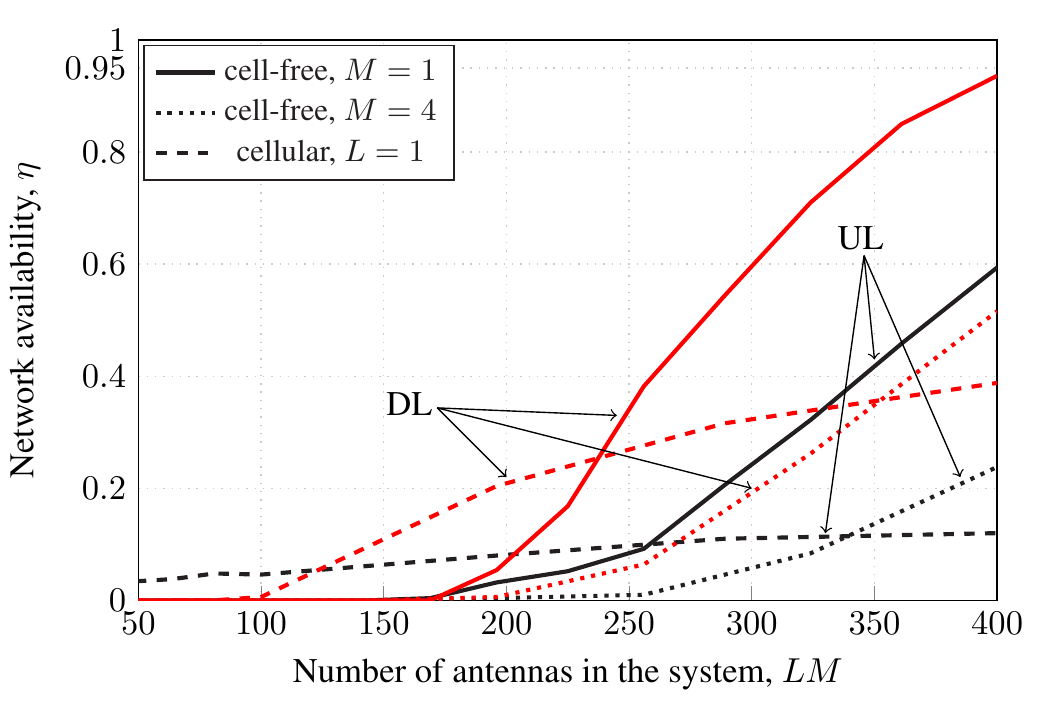}
    \caption{Network availability for $\epsilon\sub{target} = 10^{-5}$ with \gls{mr} combining/precoding in the same setup of Fig.~\ref{fig:avail_vs_antennas_MMSE}.}
    \label{fig:avail_vs_antennas_MR} 
\end{figure}

The average \gls{ul} and \gls{dl} error probabilities $\epsilon^{\mathrm{ul}}$ and $\epsilon^{\mathrm{dl}}$ for an arbitrary \gls{ue} within the coverage area are computed for fixed \glspl{ue}' positions and averaged over the small-scale fading and the additive noise. 
As performance metric, we use the \emph{network availability} $\eta$, which we define as the probability, computed with respect to the random \glspl{ue}' positions, that the error probability is below a given target $\epsilon\sub{target}$, i.e.,
 \begin{IEEEeqnarray}{rCl}\label{eq:network_avail}
 \eta = \prob{\epsilon \leq \epsilon\sub{target}}
\end{IEEEeqnarray}
where $\epsilon$ is replaced with $\epsilon^{\mathrm{ul}}$ or $\epsilon^{\mathrm{dl}}$ if UL or DL is considered. We evaluate the network availability $\eta$ for a fixed $\epsilon\sub{target}=10^{-5}$ versus the number of antennas in the system $LM$. We consider \gls{mmse} channel estimation, and both \gls{mmse} (Fig.~\ref{fig:avail_vs_antennas_MMSE}) and \gls{mr} (Fig.~\ref{fig:avail_vs_antennas_MR}) combining/precoding. 
%

For the case of \gls{mmse} channel estimation with \gls{mmse} combining/precoding (Fig.~\ref{fig:avail_vs_antennas_MMSE}), a network availability above $0.95$ in both \gls{ul} and \gls{dl} is obtained in the cell-free setting with $L=200$ single-antenna \glspl{ap} ($M=1$). 
Increasing the number of antennas per \gls{ap} does not seem to help 
  even in the \gls{dl}, where channel hardening is expected to improve as $M$ grows~\cite{chen18-06}. Indeed, although channel hardening improves, larger pathlosses due to smaller \gls{ap} densities may have a bigger impact on the \gls{dl} error probability.
For example, when $M=4$, the cell-free setting requires $L=100$ \glspl{ap} ($LM=400$) to achieve $\eta \ge 0.95$ in both \gls{ul} and \gls{dl}. 
The cellular network requires a total number of $M=400$ antennas to achieve $\eta = 0.95$. 
This illustrates the superiority of the cell-free architecture in providing \gls{urllc} services.

In all the cases considered in Fig.~\ref{fig:avail_vs_antennas_MMSE}, the \gls{dl} limits the performance. 
This is because the \glspl{ue} have no \gls{csi} and perform mismatched decoding by relying on channel hardening. 
For example, in the cell-free setting with $M=1$, the \gls{ul} requires only $L=81$ to reach $\eta =1$. However, the \gls{dl} requires $L=225$ to reach $\eta =1$. 

It is interesting to note that for both \gls{ul} and \gls{dl}, the network availability exhibits a much sharper transition from low values to high values in the cell-free setting than in the cellular setting. 
This is particularly evident for the \gls{dl}, where $\eta$ goes from around $0.05$ to around $0.95$ as $L$ is increased from $150$ to $200$. 
For the same number of antennas, $\eta$ increases from $0.46$ to $0.64$ with cellular Massive \gls{mimo}.

With \gls{mr} combining/precoding (Fig.~\ref{fig:avail_vs_antennas_MR}), a network availability close to $0.95$ can be achieved in the \gls{dl} of the cell-free setting with single-antenna \glspl{ap} only when $L=400$. 
Note that, with this number of \glspl{ap}, the \gls{dl} outperforms the \gls{ul}.
This phenomenon, which was previously noted in \cite[Sec. III-D]{ostman20-09b} in the context of cellular Massive MIMO and \gls{urllc} services, is due to two reasons:
\begin{enumerate}
  \item The number of \glspl{ap} is sufficiently large for channel hardening to occur.
  \item \gls{mr} combining/precoding maximizes the array gain without mitigating the interference, which implies that when the desired signal experiences a deep fade, the magnitude of the \gls{ul} interference is unaffected. On the contrary, when the desired signal experiences a deep fade, the \gls{dl} interference becomes small, too. This results in a larger error probability in the \gls{ul} compared to the \gls{dl}. This phenomenon, however, does not occur when \gls{mmse} combining/precoding is used. A more detailed discussion can be found in~\cite[Sec. III-D]{ostman20-09b}. 
\end{enumerate} 

Finally, it is worth highlighting that if cooperation between \glspl{ap} is not allowed in the cell-free network---a scenario referred to in~\cite{bjornson20-1a} as ``level-1 cooperation: small-cell network'', the network availability is zero when using \glspl{ap} with both $M=1$ and $M=4$ for the system parameters considered in Fig.~\ref{fig:avail_vs_antennas_MMSE} and Fig.~\ref{fig:avail_vs_antennas_MR}. 
Thus, allowing the \glspl{ap} to cooperate is crucial when deploying a cell-free network intended to support \gls{urllc}.  

\section{Conclusions}\label{sec:conclusions} 
We analyzed a cell-free Massive \gls{mimo} system supporting \gls{urllc} services, in terms of network availability.
Numerical results, based on a saddlepoint approximation of the \gls{rcus} bound (stated in Theorem~\ref{thm:rcus})  on the per-user \gls{ul} and \gls{dl} error probabilities, revealed that, in an automated-factory scenario, cell-free Massive \gls{mimo} with fully centralized processing outperforms cellular Massive \gls{mimo}.
The analysis revealed also the importance of using \gls{mmse} linear processing in place of \gls{mr} processing to obtain satisfactory performance.
Furthermore, it was illustrated that, when the \glspl{ue} have no \gls{csi} and rely on channel hardening, the \gls{dl} is typically the bottleneck from a system performance perspective, and deploying a sufficiently large number of \glspl{ap}, so as to achieve a critical \gls{ap} density, is crucial. It turned out that, for a given total number of antennas, it is more beneficial to deploy many single-antennas \glspl{ap}, than fewer multiple-antenna \glspl{ap}.

\bibliographystyle{IEEEtran}

\end{document}